\documentstyle[prl,aps]{revtex}

\input{psfig}

\begin{document}
\title{Small-Angle Excess Scattering: Glassy Freezing or Local Orientational Ordering?}

\author{H. Weber, W. Paul, W. Kob and K. Binder}
\address{Institut f\"ur Physik, Johannes--Gutenberg--Universit\"at,\\
         Staudingerweg 7, D--55099 Mainz, 
	 Germany}

\date{\today}

\maketitle

\draft

\begin{abstract}
We present Monte Carlo simulations of a dense polymer melt which shows
glass-transition-like slowing-down upon cooling, as well as a build up of 
nematic order. At small wave vectors $q$ this model system shows excess
scattering similar to that recently reported for light-scattering experiments 
on some polymeric and molecular glass-forming liquids. For our model 
system we can provide clear evidence that this excess scattering 
is due to the onset of short-range nematic order and not directly related to
the glass transition. 
\end{abstract}

The glassy freezing of supercooled fluids has been a longstanding topic of
research in a wide range of disciplines from engineering to theoretical 
physics. Despite these efforts, no agreement concerning the basic physical mechanisms
of the transition has emerged \cite{ngai,jaeckle,math1,g1}. A particularly
puzzling experimental finding is the onset of some large-scale correlations 
in various supercooled molecular and polymeric glass-forming fluids
\cite{gerharz,kanaya,fischer}. The evidence for theses correlations consists of
excess scattering at small wave numbers $q$ seen in light
scattering experiments, which by far exceeds the value to be expected from the compressibility of the system. The $q$-range corresponds
to distances of several hundreds to thousands of \AA, depending on the
material. 

This finding is completely unexpected from the point of view of theories that
treat the glass transition as a purely kinetic phenomenon, such as 
 e.g., mode coupling theory \cite{g1}. 
The only relevant length scale in this theory is the typical
inter-particle distance, that is the diameter of the ``cages'', 
the decay of which is described self-consistently by the theory.

There are, however, theories \cite{jaeckle} postulating an underlying 
second-order phase transition (possibly kinetically masked), and such theories
would imply the existence of a growing - and ultimately diverging - intrinsic
length scale associated with the glass transition, although  most existing 
theories along such lines seem questionable \cite{math1}. 
Furthermore, direct evidence for 
such a length scale from computer simulations of simple models undergoing a
glass transition is rather scarce \cite{ray,joerg,walter} and
any such lengths detected so far are quite
small and can not be connected with the large lengths seen in scattering
experiments \cite{gerharz,kanaya,fischer} far above the glass transition.

In the present work we show that this surprising scattering 
behavior of some supercooled liquids can possibly be
attributed to local orientational ordering in these systems. We will 
present a Monte Carlo simulation of the bond fluctuation model of a polymer
melt \cite{kk1,wp1} with some orientational ordering tendency.
In this lattice model each repeat unit of the chains occupies the eight vertices
of a unit cube on the three dimensional simple cubic lattice.
Each unit is connected to its neighbors along the chain with bonds fluctuating in 
length between $2$ and $\sqrt{10}$. The stochastic dynamics - single monomer hopping
or slithering snake moves - obeys connectivity and excluded volume constraints
\cite{wp1,hw1}. Using a Hamiltonian that favors long bonds, this model has 
been extensively used to study the glass 
transition in polymer melts and has been shown to reproduce much of the 
phenomenology that is observed in experiments \cite{wpjb}. 

For the present study we now employ a Hamiltonian which incorporates both bond 
length and bond angle energies \cite{hw2}. The bond length 
energy favors short bonds, $b=2$, and the bond angle energy favors
stretched angles, $\theta=180^\circ$. Details are given in \cite{hwwpkb1}.
This model system is cooled down in a stepwise fashion and is equilibrated at 
each temperature with the highly efficient slithering snake algorithm \cite{volker}. After
equilibration the dynamics of the chains is studied with the single monomer
hopping algorithm which is known to reproduce the Rouse dynamics for short 
chains in the melt \cite{wp1}. 

To study the freezing behavior we analyzed the mean square displacements of
the centers of mass of the chains in the long time limit. A measure of the
longest relaxation time of the chains can be defined as the time for which 
this mean
square displacement is equal to the mean square radius of gyration of the
chains in the melt. 
Figure $1$ shows a plot of this relaxation time, divided by its infinite 
temperature limit, for chains of length $N=20$ 
in the form of an inverse activation plot. In the 
temperature range investigated we observe a slowing down of the chain 
dynamics by two orders
of magnitude which is described well by a Vogel--Fulcher law with a Vogel--Fulcher
temperature of about $T_0=0.144$. This result would suggest the
existence of a glass transition in the melt somewhere around $T=0.2$.

In Fig. $2$ we show the scattering intensity $S(q)$  at small wave vectors.
The data are
for a system of $13732$ chains of length $N=10$ in a cubic box of linear
size $L=130$. All curves are obtained by averaging $S(q)$ over about 
$100$ statistically independent configurations. The curve at $T=0.376$ is 
flat and its value is the one to be expected 
from the compressibility of the model system \cite{wpjb}. 
Excess scattering can be found starting at temperatures around $T=0.313$ and
its intensity increases by almost a factor of $4$ upon decreasing $T$ to 
$0.219$. The $q$-range where this effect occurs 
corresponds to distances of $10-60$ lattice constants. Since one effective 
bond of our model corresponds to a group of about five successive chemical
bonds \cite{kk1,wp1,hw1,wpjb}, one estimates that the lattice spacing 
corresponds to about $2$ \AA. Thus qualitatively this
effect is similar to what was observed experimentally in certain glass-forming
materials \cite{fischer}.

In this temperature range, however, another property of our model system comes
into play: the Hamiltonian we used favors
stretched bond angles and short bonds. The conformations of the
polymer chains therefore gradually change from the Gaussian coil limit at high temperatures to the rigid rod limit at very low temperatures. In the temperature region 
where they can be considered to be almost rod-like objects they should 
therefore display a tendency for liquid crystalline ordering.  
According to Onsager theory \cite{onsager} this tendency should depend on 
the aspect ratio (length/width) of the chains in their rod-like state. 
Figure $3$ shows that we observe
exactly this behavior. At temperatures around $T \approx 0.27$ the system starts to 
develop nematic order, as shown by the behavior of the nematic order 
parameter $P_2$, which is defined as the largest eigenvalue of the Saupe
tensor \cite{sau}
\begin{equation}{
Q_{\alpha\beta} = \frac{1}{M}\left(\frac{3}{2}\left<\sum_{i=1}^M u_{i\alpha}u_{i\beta}\right> -\frac{1}{2}\delta_{\alpha\beta}\right),}
\end{equation}
where $\hat{u}_i$ is a unit vector along bond $i$ and the sum is over all bonds
in the system.
The chains of length $N=10$ show only a small increase in the value of the 
order parameter which is due to a build-up of merely local nematic order 
\cite{hwwpkb1}. The nonzero value of the order parameter for $N=10$ thus is 
just a finite size effect. 
As demonstrated below, this local ordering is, however, sufficient to produce 
the excess scattering shown in Fig. $2$ for the chains of length $N=10$. 
On the other hand the chains of length $N=20$ with their larger aspect ratio 
develop long-range nematic liquid crystalline order at temperatures below $T=0.26$ \cite{hwwpkb1}. 
For this system the excess scattering occurs in the same $q$-range and 
temperature range as for the one with $N=10$ but has
a larger amplitude. In this case the isotropic-nematic transition preempts
the glass transition, which would occur at a lower temperature.

To demonstrate the connection between the excess scattering and the 
local nematic order,
we first of all note that the temperature dependence of the
excess scattering is the same as that of the nematic short range order. To make
this link more quantitative we consider the following approximation to the
scattering function. At $T=0$ the ground state of the system consists of 
perfectly aligned rigid rods. In this case the scattering function
exactly factors into a contribution from the positional correlation
of the rods, $S_{cm}(q)$, 
and a contribution from the intra-molecular correlations, which is just
the form factor of the rods, $f(q)$:
\begin{equation}{
 S(q) = S_{cm}(q) f(q),}
\end{equation}
where the form factor of the rods is given by \cite{berne}
\begin{equation}{
f(q) = N_{scatt}\left(\frac{2}{ql} \int_0^{ql} dz \frac{\sin{z}}{z} - \left(
\frac{2}{ql} \sin{\frac{ql}{2}}\right)^2\right).}
\end{equation}
Here $N_{scatt}$ is the number of scatterers per rod and $l$ is its length.
We now approximate our scattering intensity at finite temperatures by Eq. (2),
neglecting fluctuations of the local director and the finite flexibility of
our chains.
To calculate the form factor of the rods equivalent to our chains we set 
 $l$ equal to
the end-to-end distance of the chains $\sqrt{\left<R_e^2\right>}$ and set 
$N_{scatt}$ equal to $\sqrt{\left<R_e^2\right> / \left< b^2\right>}$,
where $\sqrt{\left<b^2\right>}$ is the mean bond length.
 If we divide the scattering function for the systems with chains of length
$N=10$, presented in Fig. $2$, by the form
factor of the rods determined in this way (Fig. $4$), we see that the 
excess scattering completely vanishes. This clearly shows that the
excess scattering is due to the occurrence of orientational order in
our system and that it is even sufficient to have only 
{\em local} orientational order
to reproduce this effect.

Thus we have demonstrated that for our model system which exhibits on the one
hand 
slowing down typical for a polymer melt undergoing a thermal glass transition
and on the other hand a transition from random coil to semi-flexible and finally rigid rod
behavior of the chains and an accompanying (local) nematic ordering, the excess
scattering intensity at small wave vectors can be unambiguously ascribed to the
development of the local orientational order and {\em is not directly connected} with
the glass transition in the system. Since it is sufficient to have a tendency for
{\em local} nematic ordering for this effect to occur, the question arises
whether this mechanism may also be responsible for the excess scattering
experimentally observed for some glass-forming systems. All of these systems
are either polymers or consist of highly anisotropic molecular liquids with
the accompanying highly anisotropic interaction. While in our model the local
orientational order is associated with local chain stretching and is of
uniaxial nematic type, it is very plausible that other systems may have other
types of local orientational ordering, developing from a dense packing of 
anisotropic molecular groups. This should, however, show up in qualitatively
similar excess scattering. It would therefore be very
interesting to reexamine the local orientational properties of these
experimental systems and look at the effects of external orienting fields on the
excess scattering.

\section*{Acknowledgments}
We thank Prof. E. W. Fischer for stimulating 
discussions and gratefully acknowledge support by the EU under grant 
CIPA-CT93-0105 for one of us (H. W.).
We also thank the Sonderforschungsbereich 262 for support.

\newpage

\begin{figure}
\setlength{\unitlength}{1mm}
\begin{picture}(200,192)
\put(2,2){\psfig{figure=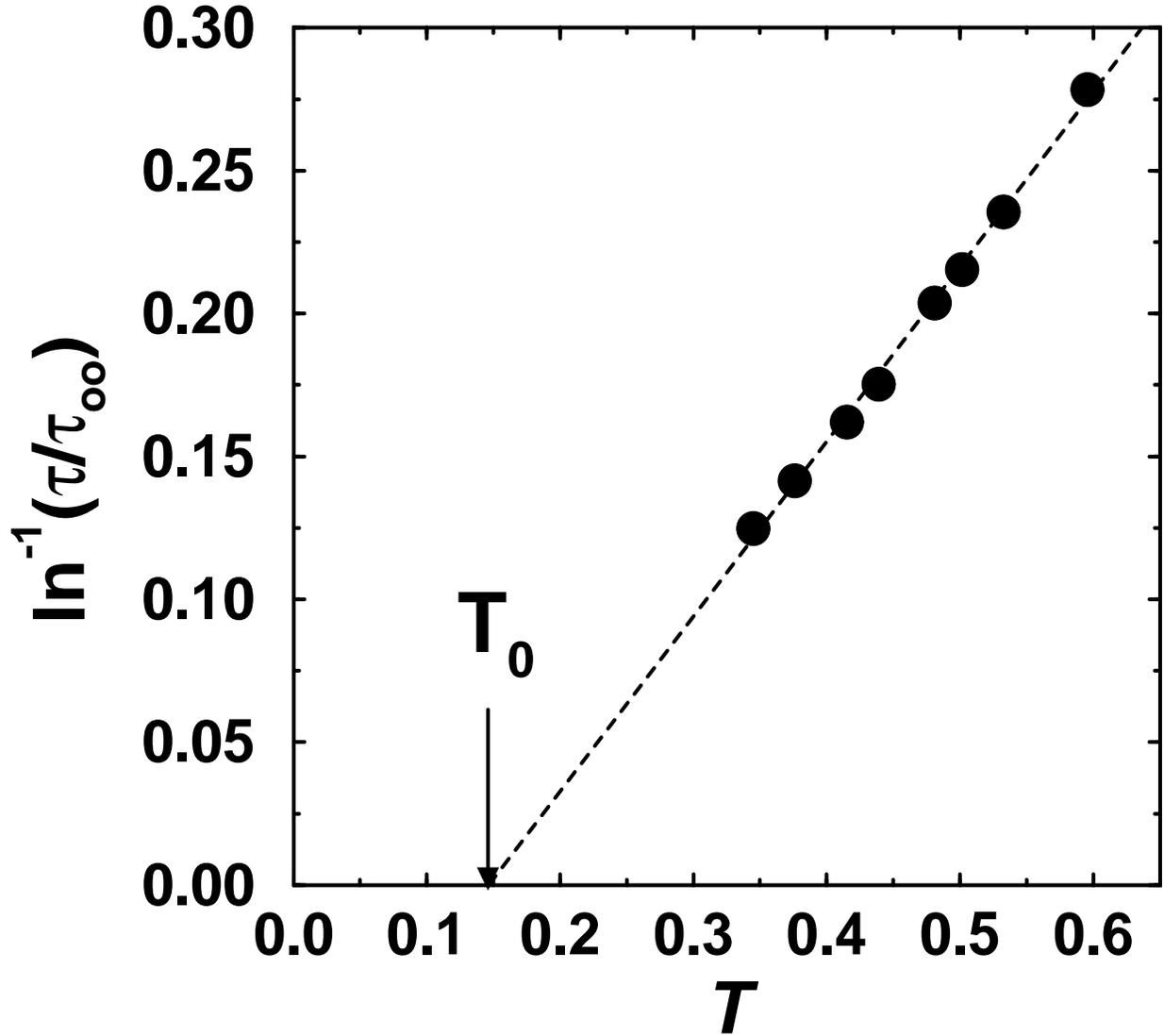,height=160mm,width=162mm}} 
\end{picture}
\caption{Longest relaxation time of chains of length $N=20$ as a function of
temperature. The dashed line shows a Vogel--Fulcher fit giving a Vogel--Fulcher temperature of $T=0.144$.}
\end{figure}

\begin{figure}
\setlength{\unitlength}{1mm}
\begin{picture}(200,192)
\put(2,2){\psfig{figure=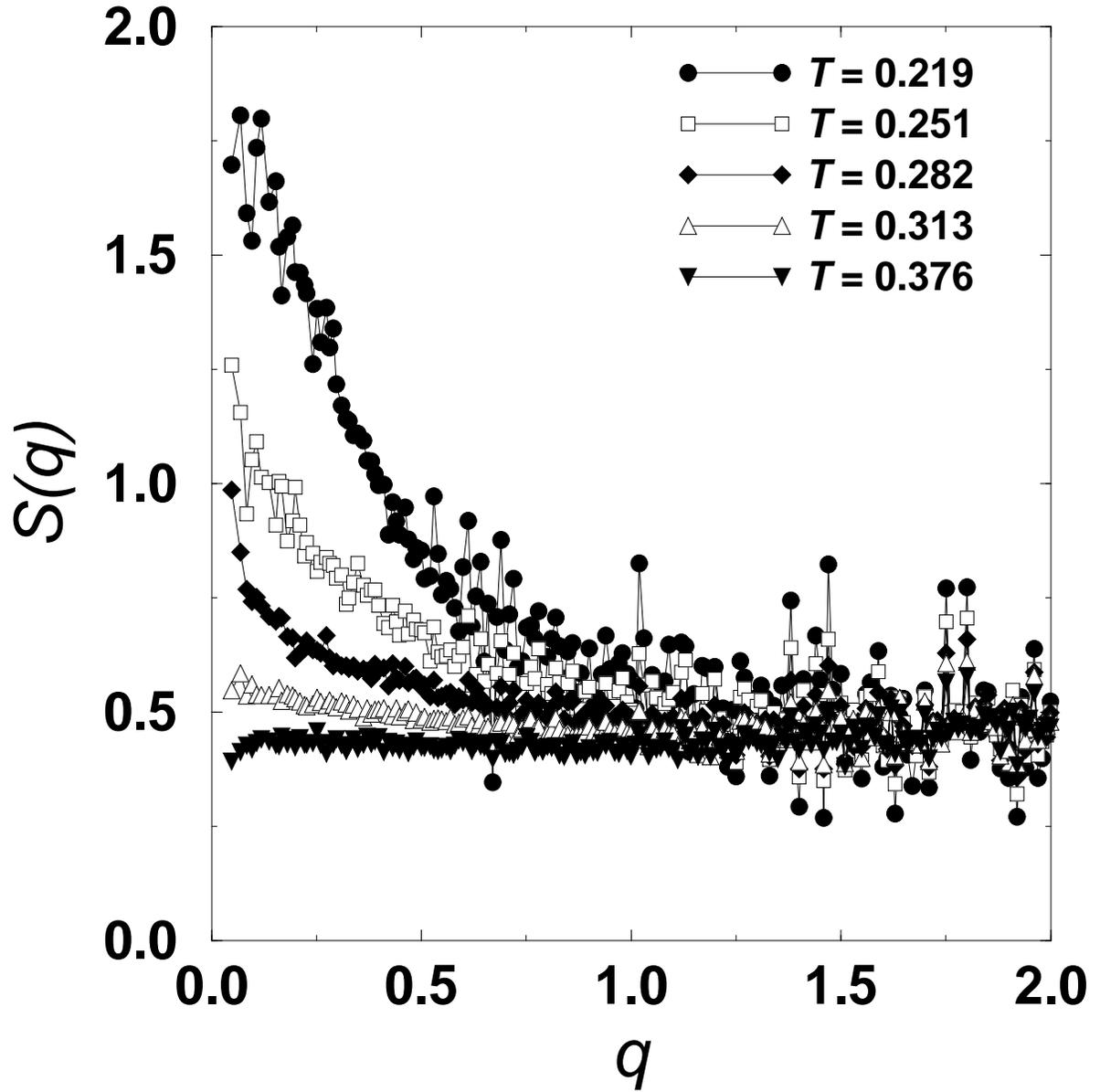,height=160mm,width=162mm}} 
\end{picture}
\caption{Scattering intensity at small wave vectors for a system of chains of length $N=10$. The different curves cover the temperature range over which one 
observes the 
build-up of local orientational order.}
\end{figure}

\begin{figure}
\setlength{\unitlength}{1mm}
\begin{picture}(200,192)
\put(2,2){\psfig{figure=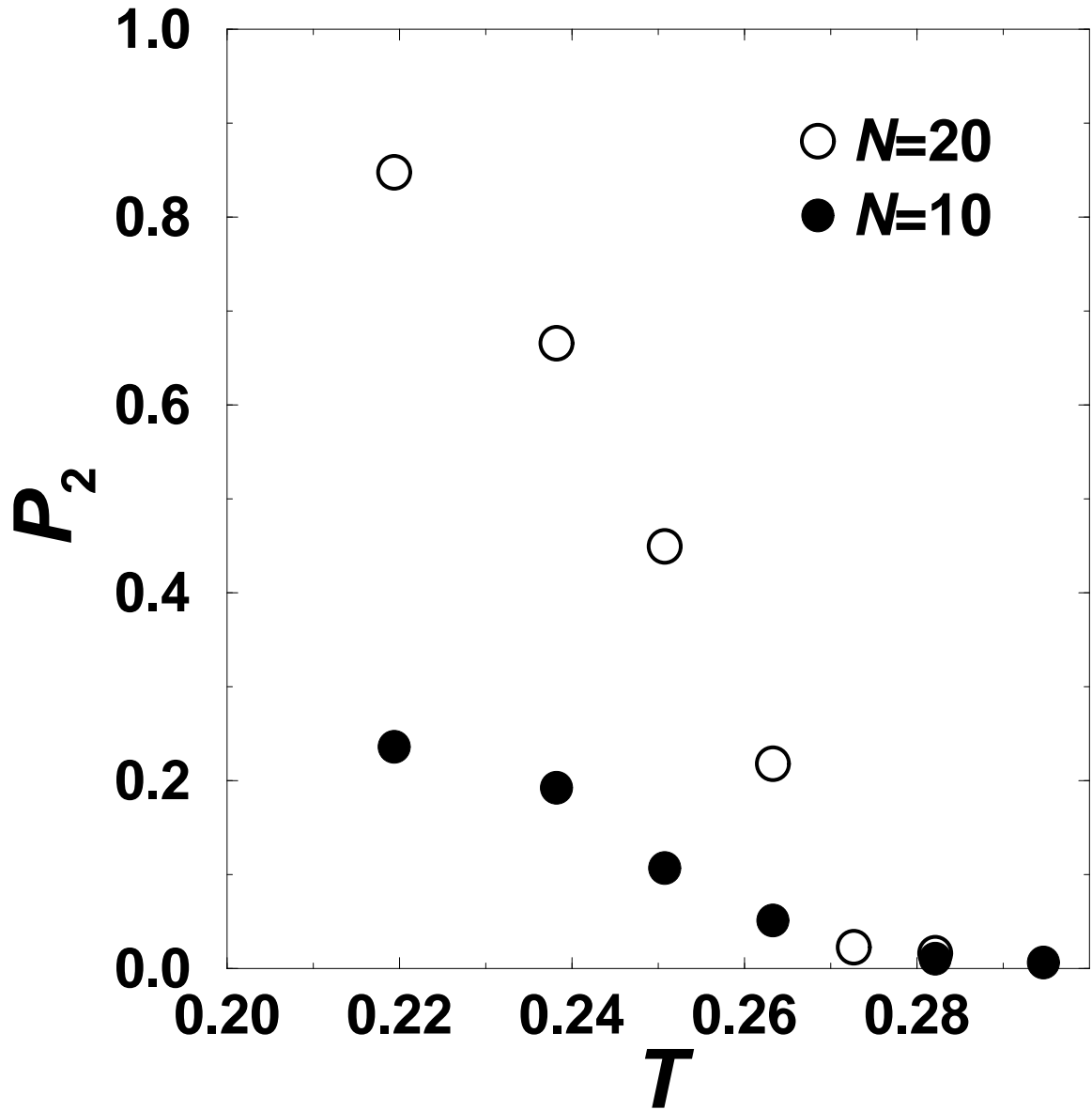,height=160mm,width=162mm}} 
\end{picture}
\caption{Nematic order parameter $P_2$ as a function of temperature. The 
open circles denote systems of $6866$ chains of length $N=20$ and the filled circles 
denote systems of $13732$ chains of length $N=10$. The linear dimension of the
box is $L=130$ in both cases.}
\end{figure}

\begin{figure}
\setlength{\unitlength}{1mm}
\begin{picture}(200,192)
\put(2,2){\psfig{figure=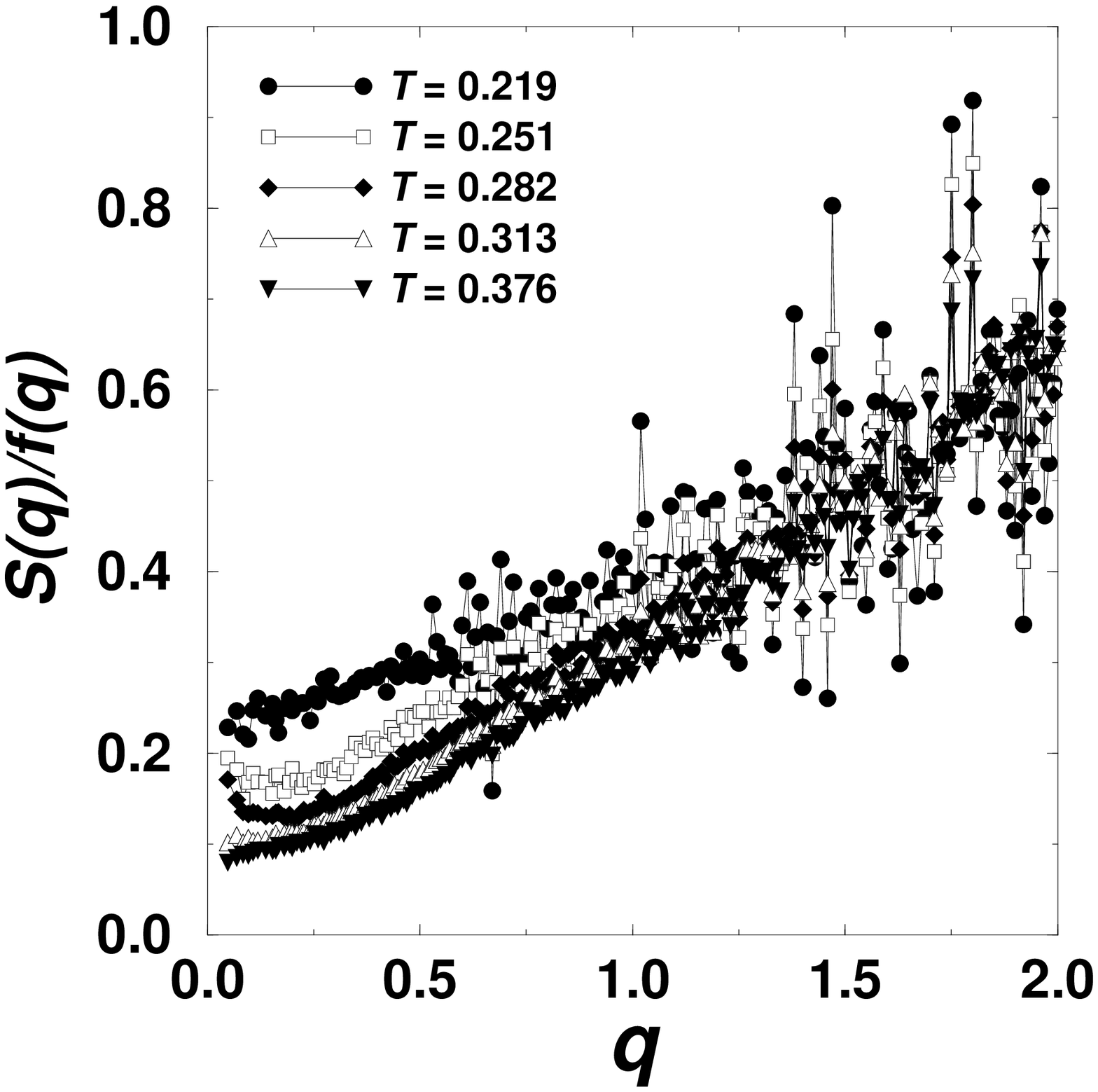,height=160mm,width=162mm}} 
\end{picture}
\caption{Scattering intensity at small wave vectors divided by an equivalent 
rigid rod form factor of the chains for different temperatures as shown in
the figure. The results are for chains of length $N=10$.}
\end{figure}


\begin{references}
\bibitem{ngai} K. L. Ngai (ed.), {\em Proc. 2nd Int. Disc. Meeting on Relaxation in Complex Systems}, J. Noncryst. Solids 172-174 (1994).
\bibitem{jaeckle} J. J\"ackle, Rep. Progr. Phys. 49, 171 (1986).
\bibitem{math1} M. Wolfgardt, J. Baschnagel, W. Paul and K. Binder, Phys. Rev. E 54, 1535 (1996).
\bibitem{g1} W. G\"otze and L. Sj\"ogren, Rep. Prog. Phys. 55, 241 (1992).
\bibitem{gerharz} B. Gerharz, G. Meier and E. W. Fischer, J. Chem. Phys. 92, 7110 (1990); E. W. Fischer, Physica 201, 183 (1993).
\bibitem{kanaya} T. Kanaya {\em et al.}, Acta Polymerica 45, 137 (1994);
Macromolecules {\bf 28}, 7831 (1995); A. Patkowski {\em et al.}, preprint (1996).
\bibitem{fischer} E. W. Fischer, E. Donth and W. Steffen, Phys. Rev. Lett. 68, 2344 (1992).
\bibitem{ray} P. Ray and K. Binder, Europhys. Lett. 27, 53 (1994).
\bibitem{joerg}J. Baschnagel and K. Binder, Macromolecules 28, 6808 (1995).
\bibitem{walter} W. Kob, in Vol. III of {\em Annual Reviews of
Computational Physics} edited by D. Stauffer, World
Scientific, Singapore, (1995), p. 1.
\bibitem{kk1} I. Carmesin and K. Kremer, Macromolecules 21, 2819 (1988).
\bibitem{wp1} H. P. Deutsch and K. Binder, J. Chem. Phys. 94, 2294 (1991), W. Paul, K. Binder, D. W. Heermann and K. Kremer, J. Phys. (France) II 1, 37 (1991).
\bibitem{hw1} H. Weber, PhD Thesis, Johannes-Gutenberg-University, Mainz, 1997.
\bibitem{wpjb} W. Paul and J. Baschnagel, in {\em Monte Carlo and Molecular
Dynamics Simulations in Polymer Science}, edited by K. Binder, Oxford University Press, New York (1995), and references therein.
\bibitem{hw2} H. Weber and W. Paul, Phys. Rev. E 54, 3999 (1996).
\bibitem{hwwpkb1} H. Weber, W. Paul and K. Binder, in preparation.
\bibitem{volker} M. Wolfgardt, J. Baschnagel and K. Binder, J. Phys. (France) II {\bf 5}, 1035 (1995); V. Tries, W. Paul, J. Baschnagel and K. Binder, J. Chem. Phys., in press.
\bibitem{onsager} L. Onsager,  Ann. N. Y. Acad. Sci. {\bf 51}, 627 (1949).
\bibitem{sau} A. Saupe, Z. Naturforsch. 19A, 161 (1964).
\bibitem{berne}B. J. Berne and R. Pecora, {\em Dynamic Light Scattering}, Krieger Publishing Company, Malabar, Florida (1990).
\end{references}
\end{document}